\documentclass[letterpaper,times]{sae}

\usepackage[dvipsnames]{xcolor}
\definecolor{SAEblue}{RGB}{1,160,233}
\raggedright

\usepackage{cite}
\usepackage[justification=justified, singlelinecheck=false]{caption}  
\DeclareCaptionFont{eightpt}{\fontsize{8pt}{11pt}\selectfont #1}
\usepackage{lineno}
\usepackage[utf8]{inputenc}
\usepackage{changes}

\usepackage{array}
\newcolumntype{L}[1]{>{\raggedright\let\newline\\\arraybackslash\hspace{0pt}}p{#1}}
\newcolumntype{C}[1]{>{\centering\let\newline\\\arraybackslash\hspace{0pt}}p{#1}}
\newcolumntype{R}[1]{>{\raggedleft\let\newline\\\arraybackslash\hspace{0pt}}p{#1}}

\usepackage{graphicx}
\usepackage{multirow}
\usepackage{amsmath}

\newcommand{\ignore}[1]{}

\usepackage[english]{babel}
\usepackage{amsthm}
\usepackage{amssymb}

\theoremstyle{definition}
\newtheorem{definition}{Definition}[section]
\newtheorem{example}{Example}
\newtheorem{problem}{Problem}

\usepackage{hyperref}
\newcommand*{\figref}[2][]{%
  \hyperref[{fig:#2}]{%
    Figure~\ref*{fig:#2}%
    \ifx\\#1\\%
    \else
      \,#1%
    \fi
  }%
}

\usepackage{nameref}
\usepackage{booktabs}

\makeatletter
\def\@seccntformat#1{%
  \expandafter\csname c@#1\endcsname\c@section
  }
\makeatother

\usepackage{subfigure}
\usepackage{multimedia}

\PaperTitle{Safety Verification and Navigation for Autonomous Vehicles 
 based on Signal Temporal Logic Constraints}
\usepackage{calc} 

\makeatletter 
\renewcommand\@biblabel[1]{#1. } 
\makeatother

\AddAuthor{Aditya Parameshwaran and Yue Wang}{Department of Mechanical Engineering, Clemson University}
\PaperNumber{2022-XX-XXXX}
\SAECopyright{2022}

\begin{document}
\maketitle
\section{Abstract}\label{s1}

The software architecture behind modern autonomous vehicles (AV) is becoming more complex steadily. Safety verification is now an imminent task prior to the large-scale deployment of such convoluted models. For safety-critical tasks in navigation, it becomes imperative to perform a verification procedure on the trajectories proposed by the planning algorithm prior to deployment. Signal Temporal Logic (STL) constraints can dictate the safety requirements for an AV. A combination of STL constraints is called a specification. A key difference between STL and other logic constraints is that STL allows us to work on continuous signals. We verify the satisfaction of the STL specifications by calculating the robustness value for each signal within the specification. Higher robustness values indicate a safer system. Model Predictive Control (MPC) is one of the most widely used methods to control the navigation of an AV, with an underlying set of state and input constraints. Our research aims to formulate and test an MPC controller, with STL specifications as constraints, that can safely navigate an AV. The primary goal of the cost function is to minimize the control inputs. STL constraints will act as an additional layer of constraints that would change based on the scenario and task on hand. We propose using sTaliro, a MATLAB-based robustness calculator for STL specifications, formulated in a receding horizon control fashion for an AV navigation task. It inputs a simplified AV state space model and a set of STL specifications, for which it constructs a closed-loop controller. We test out our controller for different test cases/scenarios and verify the safe navigation of our AV model.

\let\thefootnote\relax\footnotetext{Distribution A. Approved for public release; distribution unlimited. (OPSEC 7039)}

\section{Introduction}\label{s2}
Safety verification techniques for autonomous vehicles (AV) have seen an unprecedented rise in demand over the past few years. AVs are also becoming widely popular for off-road applications, particularly for military ground vehicles. Defining sets of rules to follow for such complex systems is an important task in maintaining the safety of the vehicle in the case of uncertainties. Temporal logic offers a formal method to help designers transform natural language rules for safety into the mathematical formulation. Over a couple of years, temporal operators have been used to express safety requirements and reach-avoid sets with the ability to time and sequence them. Some of the operators like Linear Temporal Logic (LTL)\cite{b1},\cite{b3}, Metric Temporal Logic (MTL)\cite{b4},\cite{b5}, and Signal Temporal Logic (STL)\cite{b6},\cite{b7} are used to verify and synthesize controllers for motion planning of dynamic autonomous systems that satisfy the predefined specifications. Synthesis of controllers based on LTL and MTL focus on the automata-based representation of the specification and represent the workspace as a transition system \cite{b10}. However, this leads to a highly complex transition system model that is based on discrete states that are expensive to compute and can lead to state space explosion \cite{b8}. Compared to LTL and MTL, a major advantage of STL specifications is that they provide temporal logic operators that are compatible with working with real-time continuous signals over defined periods of time.

To address concerns regarding exceeding computational time and uncertainties in the modeling, T. Wongpiromsarn et al.~\cite{b11}. have proposed using receding horizon control for temporal logic constraints problems. Wolff et al.\cite{b2} have converted the LTL and MTL specifications into mixed integer constraints to be applied recursively over a fixed horizon. Previous work has shown the effectiveness of real-time monitoring and control of STL specifications-based models for AV examples \cite{b15}. Similar applications have been more recently applied to STL specifications in the form of Model Predictive Control (MPC) problems with mixed integer constraints \cite{b13}. STL specifications assess the robustness of these time-varying parameters and provide a numerical value to verify and control specification satisfaction. As a result, robustness provides a validation measure for the STL specification even on complex multiple signals, and non-linear continuous-time systems such as AVs. Navigation tasks in AVs demand significant control over the model with assured constraints for maintaining safety. STL specifications can be converted to \textit{linear inequality equations} that can be solved recursively for any AV navigation task in a known environment over a fixed horizon \cite{b9},\cite{b16}.

 In this paper, we discuss the possibility of developing safe trajectories for our 2D AV model pertaining to minimum robustness satisfaction of our STL specifications. The STL specifications are maintained as a string user input that is initially parsed by the solver to be converted to \textit{Linear Inequality Equations} (LIE) that are solved by a Mixed Integer Linear Programmer (MILP) \cite{b14} and applied over a fixed horizon recursively in the MPC fashion. The MILP solver computes an optimal solution that minimizes a cost function based on the LIE. As MILP solves to search for a global solution, applying it for the entire state space can result in expensive computational time. Combining it with MPC has the advantage of predicting the states for only $H$ steps into the horizon and applying the first optimal control input to advance the system. The LIE are formed automatically for all sets of hard constraints such as STL specifications, system dynamics, and state and control limits of our simplified AV model. It is important to note all three listed sets of constraints are necessary conditions for the solution to be feasible. The cost function mentioned within the MILP solver provides the set of soft constraints that minimizes control input for optimally.

The rest of the paper is organized as follows. Section \ref{s3} sets up the preliminaries with an introduction to the simple car unicycle kinematic model, STL operators, grammar, and robustness. Section \ref{s4} defines the problem statement, where the MILP controller is formulated based on the set of constraints. Its application in a receding horizon-based MPC approach is also discussed. Section \ref{s5} demonstrates the simulations where the accuracy of the feedback linearized model is validated and then the MILP + MPC algorithm is set up in MATLAB and applied to solve different navigation tasks for known obstacles and goal regions. Section \ref{s6} concludes the paper and provides the scope of future work. \\

\section{Preliminaries}\label{s3}
\subsection{Simple AV Kinematic Model}
For the scope of our work, we consider a simplified 2D unicycle kinematic model for an AV  where:
\begin{equation} \label{eq1}
    \begin{aligned}
    \left[\begin{array}{c}
    \dot{x} \\
    \dot{y} \\
    \dot{\theta}\\
    \end{array}\right]=\left[\begin{array}{c}
    v \cos \theta \\
    v  \sin \theta \\
    \omega \\
    \end{array}\right] 
    \end{aligned}
\end{equation}
The states, $[x,y,\theta]$, represent the 2D pose and orientation of the AV and $[\omega, v]$ are the vehicle's angular acceleration and linear velocity, respectively. One of the major issues most nonlinear models face is the calculation of the online solution for non-convex control problems. Suppose the state space to search an optimal solution for is relatively huge. In that case, as is possible in the case of motion planning tasks for autonomous vehicles, the solver can obtain local minimum solutions that could prove to be infeasible over the trajectory. MILP problems also have longer computational times, especially when combined with multiple LIE constraints. For a simplified model as shown in Eqn. [\ref{eq1}], we can afford to linearise the system with minimal loss to accuracy. Also to avoid infeasible solutions and large computational times, we implement a feedback linearised discrete-time system model where:
\begin{equation*}
    \begin{aligned}
    &\dot{x}_1 = \dot{x} = v_x  = x_3 \\ 
    &\dot{x}_2 = \dot{y} = v_x  = x_4 \\ 
    &\dot{x}_3 = \dot{v}_x = a_x  = u_1 \\ 
    &\dot{x}_4 = \dot{v}_y = a_x  = u_2 \\ 
    \end{aligned}
\end{equation*}
where the states ${\textbf{x}}(t) \triangleq [x_1(t); x_2(t); x_3(t); x_4(t)]= [x(t);  y(t); v_x(t);  v_y(t)]$ are the global X-Y coordinates and X-Y linear velocities respectively, the control inputs are $u(t) = [u_1;u_2]$ such that $[u_1,u_2] = [acc_x,acc_y]$, i.e., the linear accelerations in X and Y coordinates, respectively. We also discretize the system with a time step $T_s = 0.1s$ as the MILP solver works only for discrete-time models. Rewriting the above equation in state space form we get:
\begin{equation}\label{eq2}
    \begin{aligned}
    &\textbf{x}(t+1) = A\textbf{x}(t) + Bu(t)\\
    &\mathbf{y}(t) = C\textbf{x}(t)
    \end{aligned}
\end{equation}
where,
\begin{equation}\label{eq3}
    \begin{aligned}
    A = \left[\begin{array}{c}
        0 \ 0 \ 1 \ 0  \\
        0 \ 0 \ 0 \ 1   \\
        0 \ 0 \ 0 \ 0   \\
        0 \ 0 \ 0 \ 0
    \end{array}\right] \ ;
 \    B = \left[\begin{array}{c}
        0 \ 0 \  \\
        0 \ 0 \   \\
        1 \ 0 \   \\
        0 \ 1 \
    \end{array}\right]
    \end{aligned} \ ;
 \  C =\left[\begin{array}{c}
        1 \ 0 \ 0 \ 0  \\
        0 \ 1 \ 0 \ 0   \\
        0 \ 0 \ 1 \ 0  \\
        0 \ 0 \ 0 \ 1   \\
    \end{array}\right] \ ;
 \  D = [0]
\end{equation}

 To compare the accuracy of the feedback linearised model with the nonlinear model (\ref{eq1}), we convert our feedback linear state outputs, $[x,y,v_x,v_y]$ into linear velocity and angular velocity, calling them $[v_{l},\omega_{l}]$. If we chose the position of the AV as the system output, we notice that:
\begin{equation*}
\left[\begin{array}{l}
\ddot{x} \\
\ddot{y}
\end{array}\right]=\left[\begin{array}{cc}
\cos (\theta) & -\sin (\theta) \\
\sin (\theta) & \cos (\theta)
\end{array}\right]\left[\begin{array}{c}
\dot{v_l} \\
v_l \omega_l
\end{array}\right]
\end{equation*}
We can then invert the matrices to calculate $[v_l,\omega_l]$ such that:
\begin{equation}\label{eq4}
\left[\begin{array}{c}
\dot{v_l} \\
v_l \omega_l
\end{array}\right]=\left[\begin{array}{cc}
\cos (\theta) & \sin (\theta) \\
-\sin (\theta) & \cos (\theta)
\end{array}\right]\left[\begin{array}{l}
u_x \\
u_y
\end{array}\right]
\end{equation}
where $[u_x,u_y] = [u_1,u_2] = [acc_x,acc_y]$ are the inputs for the second order system in the above equation. These inputs are generated as the optimal solution to the MILP problem that is presented in Section \ref{s4}. As a result, we now have the values for $[v_{l},\omega_{l}]$.In Section \ref{s5}, we compare the accuracy of our feedback linearised model velocities $[v_l,\omega_l]$ by applying them as inputs to a nonlinear kinematic model [\ref{eq1}] for a simple point-to-point navigation task and compare the results with the trajectory computed by MATLAB's Nonlinear MPC block based on the kinematics in [\ref{eq1}].
Note that within the operating domain of the states of our AV, we assume that there will always be a computed control sequence $U = [u_1,...u_{N-1}]$ such that our system dynamics $\textbf{x}(t+1) = A\textbf{x}(t) + Bu(t)$ will always be satisfied. For a control sequence of $(N-1)$ states, we have $(N)$ states on the horizon. We also assume that we can observe all our system's states at all times.

\subsection{Signal Temporal Logic (STL)}
STL has a library of operators representing the formal language's grammar and alphabet. The designers think and formulate rules in natural language that is then defined in a mathematical context using the STL operators. This allows for a more formal method to set constraints for signal-based systems.\\

\begin{definition}[STL Operators] 
Writing in the Backus-Naur form, an STL formula can be expressed with the following syntax:
\begin{equation*}
\phi ::= \top\ | \ \mu\ | \ \neg\phi \ | \ \phi \land \psi \ | \ \Box_{[a,b]} \phi\ | \ \Diamond_{[a,b]} \phi\ |\ \phi\ \mathcal{U}_{[a,b]}\ \psi   
\end{equation*}
where $\top$ is the Boolean operator true, $\mu$ is a predicate of the form $f(s_t) > c$. 
$\phi$ and $\psi$ are STL specifications and $s_t$ is a signal with respect to which we check the specifications. The other temporal operators such as $\mathcal{U}$, $\Box$, and $\Diamond$ are the `Until', `Always', and `Eventually' operators. For a given signal $s_t$ the list of temporal operators can be defined as follows: 

\begin{equation}
\begin{array}{lll}

s_{t} \models \mu & \Leftrightarrow & f\left(s_{t}\right)>0 \\
s_{t} \models \neg \phi & \Leftrightarrow & \neg\left(s_{t} \models \phi\right) \\
s_{t} \models \phi \wedge \psi & \Leftrightarrow & \left(s_{t} \models \phi\right) \wedge\left(s_{t} \models \psi\right) \\
s_{t} \models \phi \vee \psi & \Leftrightarrow & \neg(\neg\left(s_{t} \models \phi\right) \wedge \neg\left(s_{t} \models \psi\right)) \\
s_{t} \models \phi \Rightarrow \psi & \Leftrightarrow & \neg\left(s_{t} \models \phi\right) \vee\left(s_{t} \models \psi\right) \\
s_{t} \models \diamond_{[a, b]} \phi & \Leftrightarrow & \exists t^{\prime} \in[t+a, t+b] \text { s.t. } s_{t^{\prime}} \models \phi \\
s_{t} \models \square_{[a, b]} \phi & \Leftrightarrow & \forall t^{\prime} \in[t+a, t+b] \text { s.t. } s_{t^{\prime}} \models \phi \\
s_{t} \models \phi \mathcal{U}_{[a, b]} \psi & \Leftrightarrow & \exists t^{\prime} \in[t+a, t+b] \text { s.t. }\left(s_{t^{\prime}} \models \psi\right) \\
& & \wedge\left(s_{t} \models \square_{\left[0, t^{\prime}\right]} \phi\right)
\end{array}
\end{equation}

 The term $\phi \mathcal{U}_{[a, b]} \psi$ means that `eventually' in time $t' = [a,b]$  $\psi$ will be true only if from time $(0,t')$ specification $\phi$ is `always' true. The `eventually' and `always' operators mean that the specifications must be true at least once and always in the specified time range. The operators can always be defined using literals in their \textit{Negation Normal Form}. They are not restrictive as they can be applied to any of the above STL operators as shown by S. M. LaValle et al. \cite{b18}. The disjunction operator $\vee$ is an example of \textit{Negation Normal Form} using a combination of other literals. Specifications can be combined to develop constraints for autonomous vehicle models that require multiple parameters to be controlled. For every specification $\phi_i$ built on a combination of multiple state and control parameters, a single predicate value $\mu_i$ is allocated to be computed. The signal $s_t$ can be any real-time state value $\textbf{x}_{t}$ such as linear velocity ($v$), position coordinates $[x,y]$, etc.\\

\begin{example}
Consider an AV where its 2 output states are the \textit{position coordinates} $[x,y] = [x_{1};x_{2}]$. Now imagine an STL specification that calls for ($x_{1} \geq 1$) to ``always" be true for all given times. We can identify the output $y = x_1 - 1$ as the predicate $\mu_1$, such that for  $ (\mu_1 \geq 0) \ \forall \ t \ \epsilon \ [0,...\infty] $. The STL specification would be:
\begin{equation}
\phi = {\square_{[0,\infty]}
(y(t) \geq 0)}    
\end{equation}\label{eq6}
\end{example}
To justify the complete satisfaction of the above specification for all given times, we imply the method of quantifying the specification value with respect to the predicate being measured as a form of the robustness of the signal.\\
\end{definition}

\begin{definition} [Robustness Semantics for STL Specifications]
To quantitatively measure the satisfaction of an STL specification, robust semantics are used to define real-valued numbered functions $\rho^\phi$. The robustness measure for any specification $\phi$ within a system corresponding to a signal $s$ over time $t$ can be defined as:
\begin{equation}\label{eq7}
    s_t \models \phi \leftrightarrow  \rho^\phi(s_t) = \top 
\end{equation}
where the $\rho^\phi$ is defined as the robustness measure for that specification at time instant $t$, the value of the robustness measure can vary over time at every instant during the interval $t \in [a,b]$ where the specification $\phi_{[a,b]}$ holds true. The idea of the robustness measure is to provide an optimization and control parameter that can quantify the satisfaction of the specification. Positive robustness value ($\rho^\phi > 0$) guarantees the satisfaction of the specification over the time period. The robust satisfaction of $\rho(\phi \land \psi)$ is the $min (\rho^\phi$, $\rho^\psi)$. The robustness for specifications such as the "always" operator i.e. $\phi = \square \ s_{[a,b]}$ \, is \ $\rho^\phi = min_{t \in [a,b]} \ (\rho^\mu (s_t))$. Variation in the predicate value $\mu$ or in the STL specification itself would change the robustness value but vice versa is not true. Shown below is a list of the entire domain of robust semantics:
\begin{equation*}
\begin{aligned}
&\rho^\mu(s_t) \quad &=& \ f(s_t)\\
&\rho \neg \mu(s_t) \quad &=& \ -\mu(s_t)\\
&\rho^{\phi \wedge \psi}(s_t) \quad &=& \ \min \left(\rho^{\phi}(s_t), \rho^\psi(s_t)\right)\\
&\rho^{\phi \vee \psi}(s_t) \quad &=& \ \max \left(\rho^{\varphi}(s_t), \rho^\psi(s_t)\right)\\
&\rho^{\square_{[a, b]} \phi}(s_t) \quad &=& \ \min _{t^{\prime} \in[t+a, t+b]} \rho^{\phi}\left(s_{t^{\prime}}\right)\\
&\rho^{\phi} \mathcal{U}_{[a, b] \psi}(s_t) &=& \ \max _{t^{\prime} \in[t+a, t+b]}\left(\operatorname { m i n } \left(\rho^\psi\left(s_{t^{\prime}}\right)\right.\right. \text , \left.\min _{t^{\prime \prime} \in\left[t, t^{\prime}\right]} \rho^{\phi}\left(s_{t^{\prime \prime}}\right)\right)  
\end{aligned}
\end{equation*}
\end{definition}

\begin{example}(Specification Example for Navigation Task): \\
An example of STL specification for an AV to avoid an obstacle with known bounds ($P_{obstacle}$) and end at a goal location with known bounds ($P_{goal}$) within the time frame $[a,b]$ would look as follows:
\begin{equation}\label{eq8}
    \phi = \square\neg P_{obstacle} \land \diamond_{[a,b]}P_{goal}
\end{equation}
\end{example}
Here, $P_{obstacle}$ and $P_{goal}$ are convex polytopes developed based on the X-Y coordinates provided by the user as inputs in the STL specification. For the navigation task, multiple STL formulae would be combined to solve for an optimal solution such that $\phi' = (\phi_1 \ \land \phi_2 \ \land ...,\phi_{n})$.
We use an STL parser to decode the string-based specification into LIE that are then encoded as the bounds for the obstacle regions and goal regions into the MILP problem. The MILP problem is then solved recursively for a fixed prediction horizon ($H$) using a receding horizon MPC.

\section{Problem Statement}\label{s4}
The problem at hand is to parse the STL specifications set for the system and bound them as hard constraints to an MPC-based controller that minimizes a cost function. In general, the signals ($s_t$) for STL specifications can be any of the states ($\textbf{x}$), control inputs ($u$) or the state outputs ($y$). However, it is to be noted that for the problem at hand, we work with specifications that demand control only of the system states ($\textbf{x}$), such that $ s_t \in \textbf{x}(t)$. Moving on, we shall replace the signals $s_t$ with the state values $\textbf{x}(t)$ for easier understanding of the problem formulation. The MPC controller solves for an optimal control sequence $u_{i} = [u_1,...u_H]$ within the receding horizon of $H$ steps while satisfying the hard constraints set by STL, system dynamics and state constraints.\\
\begin{problem} (Cost Function Synthesis from STL): \\
For an initial value of $x_0$, prediction horizon of $H$ steps going from time $t_0,t_1,...t_H$, cost function $J$ and STL formula $\phi$, we calculate for the optimal control at each time step as:
\begin{equation}\label{eq9}
\begin{aligned}
    & min_u^{H,t} J(\textbf{x}(t),u^{H,t}) \\
    & \text { s.t. } \ \rho^\phi(\textbf{x}(t)) \geq \rho^{\phi}_{min} \\
    & \textbf{x}(t+1)= A\textbf{x}(t) + Bu(t) \\
    & u_{t} \in \mathcal{U}(t) \ , \ \textbf{x} \in \mathcal{X}(t)
\end{aligned}
\end{equation}

where $u^{H,t}$ is the control input to be calculated for each time step $t$ as defined by V. Raman et al. \cite{b15}. The robustness measure $\rho^\phi(\textbf{x}(t))$ for the specification $\phi$ based on state values $\textbf{x}$ at time $t$ should be greater than a minimum robustness value of $\rho_{min}$
\end{problem}

\begin{problem}(Encoding STL specifications into MPC): \\
The task of the controller is to compute for an optimal solution such that the STL specif,icatconstraints $\phi$ maintains a positive robustness value $\rho^\phi({\textbf{x}}(t)) > 0$ for the state values $\textbf{x}$ within the horizon length $H$. The specification for the horizon can be computed repeatedly as shown by A. Dokhanchi et al.\cite{b17}:
\begin{equation}\label{eq10}
\begin{aligned}
&h^\mu=0, \\
&h^{\neg \varphi}=h^{\varphi}, \\
&h^{\vartheta_{[a, b] \varphi}}=h^{\square_{[a, b]} \varphi}=b+h^{\varphi}, \\
&h^{\varphi \wedge \psi}=h^{\varphi \vee \psi}=\max \left(h^{\varphi}, h^\psi\right), \\
&h^{\varphi} \mathcal{U}_{[a, b]} \psi=b+\max \left(h^{\varphi}, h^\psi\right),
\end{aligned}
\end{equation}
where  $\mu$ is the numerical predicate with $\phi, \psi$ being the STL specifications. For the predicate horizon length of $h \leq H$, we can calculate the robustness value $\rho^\phi(t) \forall \ t \ \in \ (0,...h)$  for the signal $\textbf{x} \in (\textbf{x}_t,\textbf{x}_{t+1},\cdots,\textbf{x}_{t+h})$. This is possible because we predict state up to $H$ steps but only calculate the robustness values of the states up to $h$ steps. We know the current robustness value at time step $t$ at our current state $\textbf{x}(t)$ and also know the history of all the values leading to the current state. Based on the evolution of our system, we try to maintain a positive robustness for all future predicted states from $\textbf{x}_t,\cdots, \textbf{x}_{t+h}$. As we are working with an MPC controller, the most distant robustness value we can calculate would be based on the prediction horizon $H$ for the controller, i.e., $\rho^\phi(t+H)$. Hence the set of constraints that we set range from the previous known horizon length ($t-h$) till the future predicted state at prediction horizon $t+H$. This means that we try to find an optimal solution that would satisfy the positive robustness measure for all the steps within the prediction horizon as described by Sadraddini et. al \cite{b19}: 
\begin{equation}\label{eq11}
\begin{aligned}
\left\{\begin{array}{cc}
\rho(s_t)^{\phi}\left[t-h\right] & \geq 0 \\
\rho(s_t)^{\phi}\left[t-h+1\right] & \geq 0 \\
\vdots\\
\rho(s_t)^{\phi}\left[t+H\right] & \geq 0
\end{array}\right.
\end{aligned}
\end{equation}
The calculated control sequence would be $u^{H,t}(t) = [u(t),u(t+1),$ $\cdots,u(t+h+H)]$. Only the first control input $u_t$, i.e., at the first step, is applied to control the system. The prediction process starts again, along with calculating new robustness values based on the new current state for $t = [t+1,...t+H+1]$.
\end{problem}

 The main aim of our controller is to compute correct and optimal solutions. Even though optimal solutions are ideally sought after, suboptimal solutions that satisfy the constraints are also accepted. The set of hard constraints we define for our problem includes the state and control $[\mathcal{X},\mathcal{U}]$ limits, the system dynamics for our AV model ($\textbf{x}(t+1) = A\textbf{x} + Bu$) and the set of STL specifications $\phi'$. 

\subsection{Mixed Integer Linear Programming}

In this section, we formulate the MILP problem based on the state, control and STL constraints for our system. A general MILP based for our system would be in the form of: 
\begin{equation}\label{eq12}
\begin{aligned}
    &\min _{u(t)} \sum_{k=0}^{H-1}\left\|u(t+k)\right\| \\
    & \textbf{G}\textbf{x}(t) \geq \textbf{l} \quad  \\
    & \text { s.t. } \ \rho^\phi(\textbf{x}(t)) \geq \rho^{\phi}_{min}, \\
    & \textbf{x}(t+1)= A\textbf{x}(t) + Bu(t), \\
    & u(t) \in \mathcal{U}(t) \ , \ \textbf{x}(t) \in \mathcal{X}(t)
\end{aligned}
\end{equation}
where $\min _{u(t)} \sum_{k=0}^{H-1} ||u(t+k)||$ denotes the cost function to be minimised subject to \textit{Linear Inequality Equations} (LIE) $(\textbf{G}\textbf{x}(t) \geq \textbf{l})$ that are formulated using system dynamics $\textbf{x}(t+1)= A\textbf{x}(t) + Bu(t)$, control and state constraints $u(t) \in \mathcal{U}(t) \ , \ \textbf{x}(t) \in \mathcal{X}(t)$ and the robustness value for the trajectory $\rho^\phi(\textbf{x}(t)) \geq \rho^{\phi}_{min}$. The optimal solution for the problem would be the one that satisfies all these constraints and minimizes the cost function. As safety is a key feature for our model, we ensure that these constraints act as a sufficient and necessary condition to compute the optimal solution. The aim of the solver is to calculate a feasible trajectory $(p)$ for every prediction horizon $H$ as discussed by R. Koymans et al.\cite{b4}. The robustness measure $\rho^\phi(p)$ of the calculated trajectory determines whether the trajectory sustains the constraints or voids it. The set of STL constraints are converted in LIE as discussed later in this section. Although satisfaction of these constraints depends upon the robustness of the specification $\phi'(t)$ at any time $t$, the safety of the vehicle depends only upon the specifications related to the unsafe regions i.e. $\rho^{\phi_{unsafe}}(t) \geq 0$ at all times.

Using the toolbox provided by \cite{b7}, which is based on {\fontfamily{cmtt}\selectfont
sTaLiRo}, we calculate the critical robustness, time and predicate for the infeasible trajectory. As we already know the bounds for the obstacle region $P_{obstacle}$ and the state space bounds, we can represent the points in the safe region as a polytope $P_{safe} = (P_{state} - P_{obstacle})$. Any such set of polyhedrons, with $f$ faces can be represented by the below shown \textbf{LIE}:
\begin{equation}\label{eq13}
P = \{ \textbf{x} \ | G\textbf{x} \leq l,\ G  \in  \mathbb{R}^{f x n},\ l \in \mathbb{R}^f \}  
\end{equation}
If the STL specification $\phi$ needs to be true at any time $k$, i.e. the trajectory  $p^\phi$ needs to be in the set $P(k)$, then we represent the LIE as:
\begin{equation*}
    G\mathrm{x(k)} \leq l
\end{equation*}
To counter this, if we require the STL specification to not be true at the time $k$, then we make use of a binary encoding method, where for any specification $\phi$, a binary variable $z^{\phi}(k) \ \in \ \{0,1\}$ indicates whether $\phi$ stands true or false at the time $k$. For $z^{\phi}(k) = 1$, the specification stands true and for $z^{\phi}(k) = 0$, the specification stands false. We also use a very large number $M$ such that the overall LIE look like:
\begin{equation}\label{eq14}
    \begin{aligned}
    G\textbf{x}(k) + Mz^{\phi}(k)    \geq l , \ z \ \in \ \{0,1\}^f
    \end{aligned}
\end{equation}
For instance,if the calculated trajectory is as required (i.e. $\rho(p,k) \ \geq \ \rho_{min}$) and the predicate $\mu(k)$ is within the $P_{safe}$ region, then $z=1$ and the LIE constraints are simply to take the AV to the goal region as $M$ is sufficiently large to always the LIE holds true for any $\textbf{x}(k)$. But if $\rho(p,k) \ \geq \ \rho_{min}$  at time $k$, then $z=0$ and the obstacle region LIE are also enabled, thus forcing the MILP solver to calculate a new safe trajectory $p$ from $t = [k,....k+H]$. This method helps reduce computational time as we only enable the unsafe region constraints when our predicted states have a robustness value lesser than the minimum required. This is still a safe controller as any positive robustness value ensures that the AV does not collide with the obstacle. Generally as shown in example \{\ref{ex4}\}, we typically have multiple sets of STL specifications combined to form $\phi'$ usually separated by the $\land$ operator indicating that all of them have to be satisfied at all given times (or the time range $[t_a,t_b]$ in which they are specified). The STL parser developed separates the specifications and parses them individually to compute the polytope (G,b). For instance, for the AV navigation task, the typical STL list would look like $\phi' = (\phi_{goal} \ \land \ \neg\phi_{obstacle})$ which would be separated into sets of LIE.

\begin{definition}(Combining Multiple Linear Inequality Equations)
Let us call an LIE ($G,l$), defining a single polytopic region ($P$) in space, as $L$ where $L = l - G\textbf{x}$. To maintain a minimum robustness of $\rho^{\phi}_{min}$, we modify the LIE such that
\begin{equation}
\begin{aligned}
    & L^{\phi} =  b \pm \rho^{\phi}_{min}*1_f - G\textbf{x}, \\
    & L^{\phi_{obstacle}} = - b \pm \rho^{\phi_{obstacle}}_{min}*1_f + G\textbf{x} + Mz^{\phi}
\end{aligned}
\end{equation}

where $1_f$ is a column vector of ones of the size of $f$. We can stack multiple such modified LIE's $L$ for each obstacle or goal region within the workspace such that $L_{goals} = [L_{goal^1},...L_{goal^{g}}]$ for ($g$) goal regions and $L_{obstacles} = [L_{obstacle^1},...L_{obstacle^{o}}]$ for ($o$) obstacle regions. For any trajectory $p$ within the horizon $H$, the combined LIE set $\hat{L}$ would be:
\begin{equation}
\begin{aligned}
  \hat{L} = [\alpha_{(g,k)}*L_{goal_{i}} \ \land \ L_{obstacle_{j}} \ \land \\ L_{state}\ \land \ L_{control}\ \land \ L_{dynamics}]
\end{aligned}
\end{equation}
where $\alpha(g,k)$ is a column vector with binary parameters of $g$ rows that dictate which one of the goal LIE's would be active for all time $k = [0,1,...N]$.  For instance, when the AV needs to go to $goal_1$ at time $k$, then  $\alpha(1,k) = 1$, and $\alpha(2:g,k) = 0$ would be inactive. This creates a hierarchy within the waypoints defining an order in which the waypoints would be visited. As a result, this reduces the size of the overall set of constraints, reducing the MILP solver computational time. 
\end{definition}

\subsection{STL-based Receding Model Predictive Control}
In this section, we discuss the MPC solution to the MILP problem discussed previously based on the list of constraints. As mentioned earlier, a trajectory $p(t_0) = p_0$ for the first time step $t_0$ is calculated by ignoring any $\phi_{obstacle}$ and only applying constraints based on the state, control, system dynamics and remaining STL for a prediction horizon $H$. If the initial state of the system $x_0$ is within the bounds, and the trajectory calculated is a feasible one, then the control input $u(t_1)$ is applied for the first time step, and the prediction process is started again for time $t = (1,..H+1)$. \\
\ For any iteration in the future, if the predicate value $\mu(p_k) \leq \rho^\phi_{min}$ for the trajectory $(p_k)$, then the calculated trajectory is infeasible at time step $(k)$. This is identified by calculating the critical robustness value for the trajectory using {\fontfamily{cmtt}\selectfont sTaLiRo}. If the trajectory is infeasible and within an obstacle region, then the binary variable is set as $z = 0$, and the STL constraints regarding the obstacle region are set as an additional set of constraints for the MILP problem to solve again for the time range $t = t+k,...t+k+H$. This occurs recursively till the trajectory is a feasible one and only then do we apply the control input $u(k)$ to advance the system by one step. The cost function that we are minimising for our problem is:
\begin{equation}\label{eq15}
\begin{aligned}
&\min _{\vec{u}_t} \sum_{k=0}^{H-1}\left\|u_{t+k}\right\|  \\
&\text { s.t. } \ \textbf{x}(t+k+1)= A\textbf{x}(t+k) + Bu(t+k) \\
&\textbf{x}(t) \models \phi \text { and } u(t+k) \in \mathcal{U}(t+k), \quad k= (0, \ldots, H-1) \
\end{aligned}
\end{equation}

where we are minimizing the norm of the control input $u_{k,..k+H}$ subject to a set of mixed-integer constraints, system dynamics with state, and control bounds. The control input finally applied to propagate the system is $u_k$. We first apply the state, control, and system dynamics constraints to calculate a feasible trajectory from the start point $P_{start}$ to the goal location $P_{goal}$. This means that even though the set of STL constraints $\phi'$ are not first applied during the MILP synthesis, we check the robustness value of the trajectory $\rho^{\phi'}$. If the initial trajectory has $\rho^{\phi'}\ \geq \ \rho_{min}$ for time range $(t = 0,..., H)$, then the computed control values are applied in a receding horizon fashion. As discussed in equation (11), we calculate the robustness values for time period $(t-h,..t+H)$, and if any $\rho^{\phi'}(t) < 0$ then we term it as an "infeasible trajectory".\\

\subsection{Specification Parser for Navigation Tasks}
The STL specifications mentioned for each navigation task are in the form of string-based inputs that need to be decoded to extract useful information for our MILP formulation. \\
\begin{definition} (STL Parser)\\
A custom string parser is developed specifically to understand the STL-specification inputs. As defined in the MILP section, the idea of the parser is to identify the 2D coordinates for the obstacle, safe, and goal regions from the STL specification. Those coordinates are then converted into convex polytopes that can be defined in 2D space with LIE. The LIE's then act as a set of hard constraints for the MILP to solve for an optimal solution. The constraints based on the specification followed by the $``alw_{[a,b]}"$ operator are recursively applied at every time step throughout the time period $[a,b]$. For the $``ev_{[a,b]}"$, the constraints are applied throughout the period $[a,b]$, but once they are satisfied they are discarded from the specification list.
An example STL specification input looks as shown: \\
\end{definition}
 
\begin{example}\label{ex3}(\textit{Go-To-Goal}):\\
\begin{equation*}
\begin{aligned}
``& alw (not(X(1:2,t) \in [x_{obstacle},y_{obstacle}]) \ and \\
 & ev_{[a,b]}(X(1:2,t) \in [x_{goal},y_{goal}])"
\end{aligned}
\end{equation*}
where $x_{obstacle} \in [\textbf{x}_1,\textbf{x}_2,..\textbf{x}_r]$ and $y_{obstacle} \in [\mathbf{y}_1,\mathbf{y}_2,..\mathbf{y}_r]$ are list of X-Y coordinates for the obstacle region, ``$X(1:2,t)$" are the AV's X-Y position values at time $t$ and $[t+a,t+b]$ is the range of time, starting from $t=0$, within which we need our AV to be at the goal location. The above example only deals with a single obstacle region. For multiple obstacle regions, where the STL specification lists looks $[P_{obstacle_1},...P_{obstacle_n}] \Rightarrow [\phi_1,...\phi_n]$, the combined specification list for the avoiding the obstacle regions would be $\square(\neg\phi_1 \ \land \ \neg\phi_2 \ ... \land \ \neg\phi_n)$ \\
\end{example}
\begin{example}\label{ex4}(\textit{Multi-Waypoint Navigation}): \\
\begin{equation*}
\begin{aligned}
``& alw (not(X(1:2,t) \in [x_{obstacle},y_{obstacle}]) \ and \\
& ev_{[a_1,b_1]}(X(1:2,t) \in [x_{goal_1},y_{goal_1}]) \ and \ ... \\
& ev_{[a_{n},b_{n}]}(X(1:2,t) \in [x_{goal_n},y_{goal_n}])"
\end{aligned}
\end{equation*}
where we consider the set of goal coordinates to be $[(x_{goal_1},y_{goal_1}),...(x_{goal_n},y_{goal_n)}]$. The time stamps related to the entry and exit times for each of the way-points are also set as $[a_i, b_i] \Rightarrow \phi_{goal_i} \forall i=1,..n$. It is important to note that the time stamps are required to be generally in the range in which we expect the AV to reach the waypoint. Mentioning coinciding time ranges would lead to infeasible solutions. The solver will return an infeasible solution if the AV cannot reach the waypoint within the time range provided. Hence it is best to identify large ranges are inputs that can guarantee convergence.
\end{example}

\section{Simulations}\label{s5}
To test the validity of our model, we carry out simulations on MATLAB 2022B using {\fontfamily{cmtt}\selectfont Gurobi} solver through {\fontfamily{cmtt}\selectfont YALMIP} for MILP formulation on a \textit{i9 8th gen} processor with \textit{2.9GHz} clock speed and \textit{16GB} of RAM memory. The model is initialised for a sampling period $T_s = 0.1s$, with initial states $\textbf{x}_0 = [-0.6,-0.4,0,0]$ for \textit{Go-To-Goal} and $\textbf{x}_0 = [0,-0.4,0,0]$ for \textit{Multi-Waypoint} with a maximum run-time simulation period of 100 seconds. If the AV reaches its goal target within the simulation period, then the simulation ends.

\subsection{Linear vs Nonlinear Model Comparison}

\begin{figure}[!ht]
    \centering
    \includegraphics[scale = 0.15]{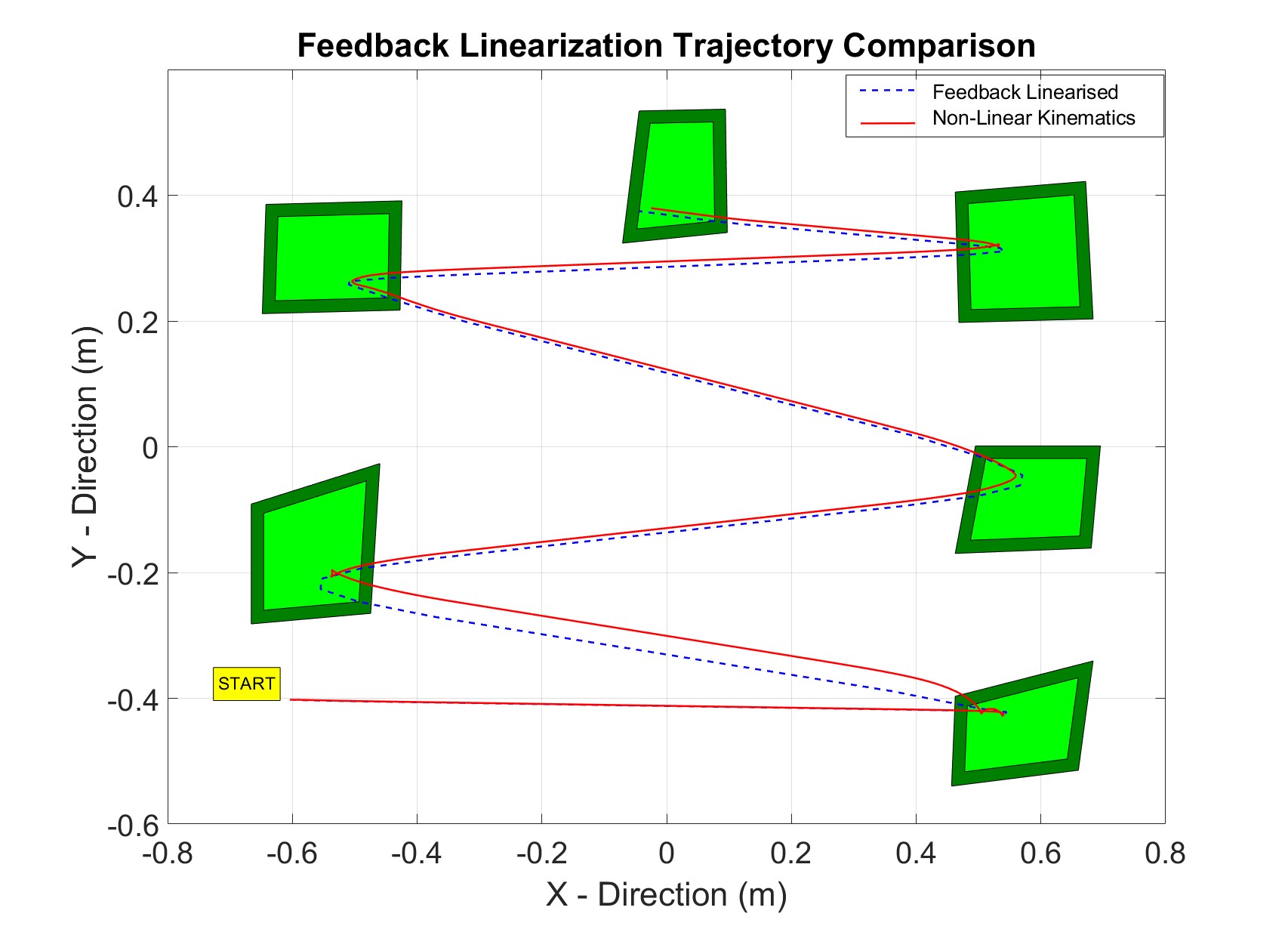}
    \caption{Comparison between Linear and Non-Linear system models}
    \label{fig1}
\end{figure}

As we are using a feedback linearised model with the state and input matrices as shown in [\ref{eq3}], instead of the non-linear AV system described in [\ref{eq1}], we will first compare the results between the two systems to test the accuracy of the linear discrete-time model. For the non-linear system, we use the ``\textit{Nonlinear MPC}" (NMPC) block provided in MATLAB with the same simulation time-step $T_s = 0.1s$, the same as used to discretize the linear system as well. The cost function to minimize for the NMPC model is:
\begin{equation}\label{eq16}
\begin{aligned}
&\min _{\vec{u}} \sum_{0}^{H-1} \left\|\textbf{x}^{ref}-\textbf{x}\right\| \ +\ \left\|u\right\| \\
&\text { s.t. } \ \dot{\textbf{x}}= f(\textbf{x}(t),u(t)) \\
& u \in \mathcal{U}, \qquad k=0, \ldots, H-1
\end{aligned}
\end{equation}

where, $\dot{\textbf{x}}= f(\textbf{x}(t),u(t))$ is the system dynamics provided in [\ref{eq1}] for NMPC and in [\ref{eq2}] for Linear MPC. A reference trajectory ($\mathrm{x^{ref}}$) is provided to both models, that is, a set of points between the start and goal locations. The cost function formulation differs from the one we used for the MILP + MPC model, as the main task of that system was to compute and traverse a safe trajectory. It is important to note that we are comparing the two systems to verify the accuracy of our feedback linearized state space model. Once we simulate the system, the states of our non-linear model are $[\textbf{x},\mathbf{y},\theta]$ that can be easily plotted. The simplified AV kinematic model requires $[v,\omega]$ i.e., linear velocity and angular velocity as inputs. Using the linear and angular velocities developed from the linear feedback model [\ref{eq4}], we simulate the non-linear simplified kinematics AV model described in (\ref{eq1}) and verify the accuracy of our linearisation. By observing (\ref{fig1}), we note that the trajectories for discrete-time feedback linearized and the continuous time non-linear models have a small deviation through the 6 waypoints which shows a good approximation. The green regions indicate acceptable waypoints for the AV to reach.

\subsection{STL Constraint-based Simulations}
\begin{figure}[!ht]
    \centering
    \includegraphics[scale = 0.45,width = 90mm]{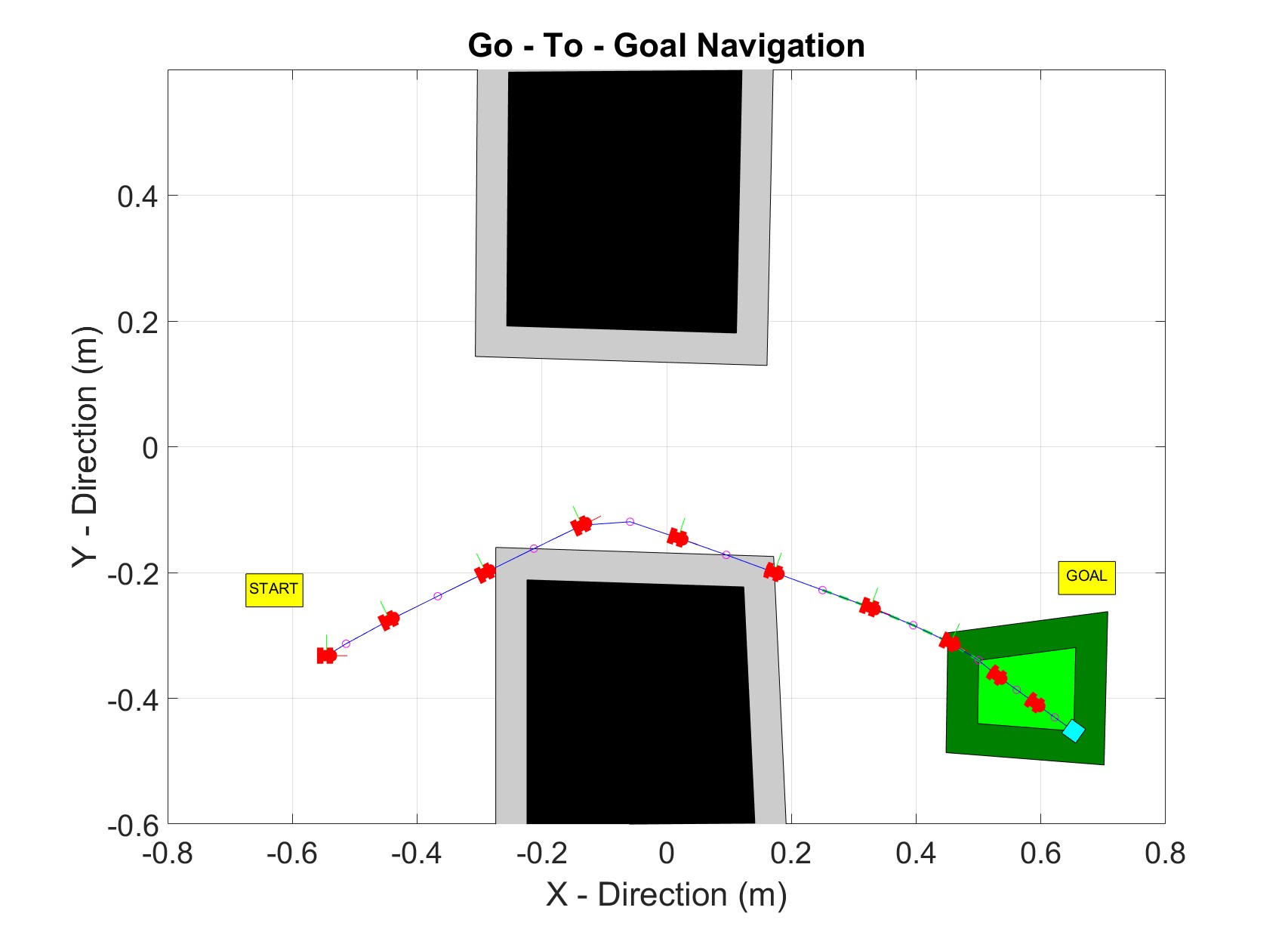}
    \caption{\textit{An example of the Go-To-Goal STL specification}}
    \label{fig2}
\end{figure}

\begin{figure}[!ht]
    \centering
    \includegraphics[scale = 0.45,width=90mm]{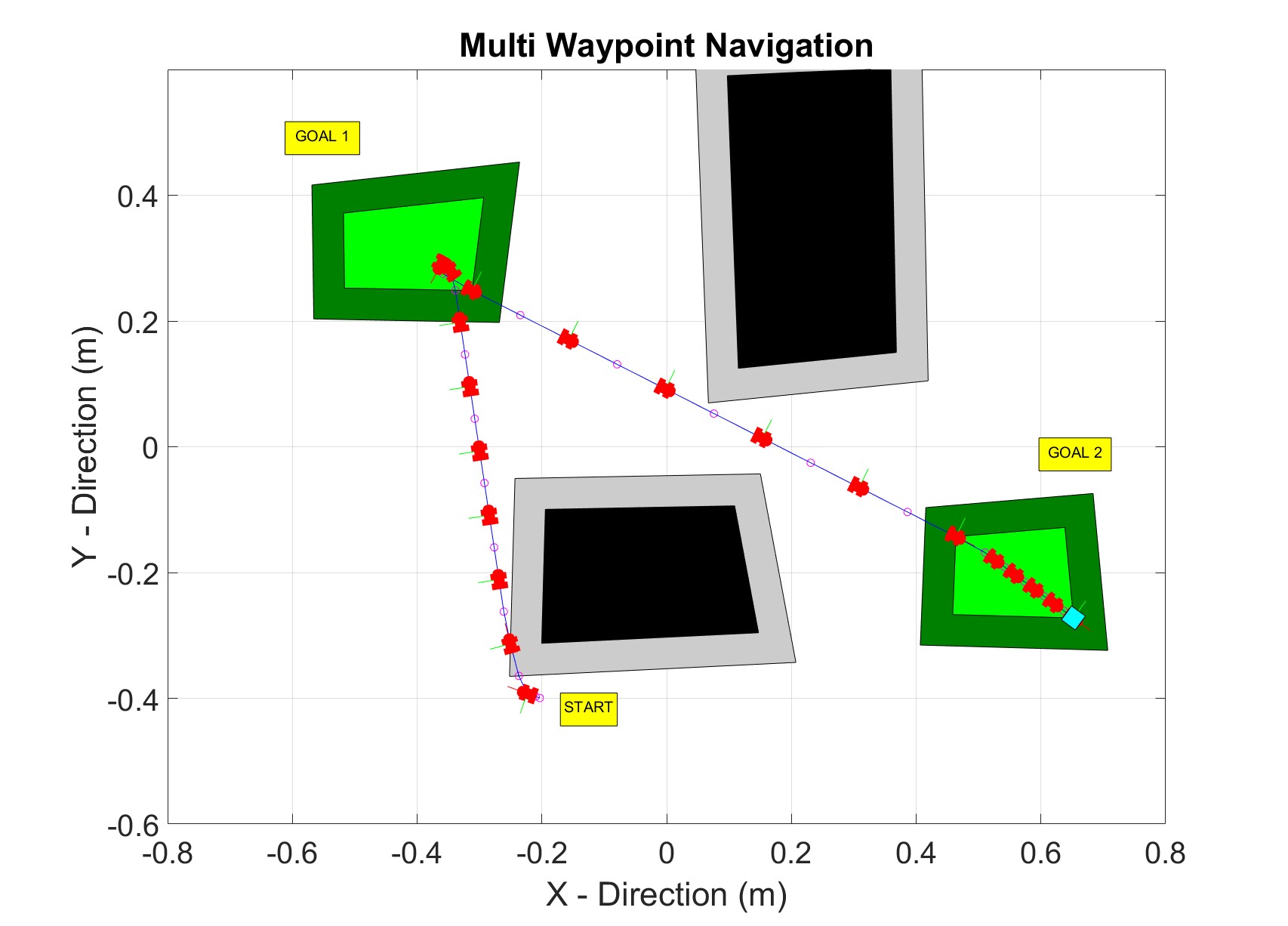}
    \caption{\textit{An example of the Multi-Waypoint STL specification}}
    \label{fig3}
\end{figure}

 Now we test our linear discrete-time model for the two examples, \{\ref{ex3}\} and \{\ref{ex4}\} mentioned in the previous section. We chose a prediction horizon $H \ = \ h = 20$ steps where $H$ is the prediction horizon, and $h$ is the predicate horizon. The predicate horizon captures the trajectory's future robustness values for $h$ steps. The state bounds are $|\mathbf{x(1:2)}| \leq 1m$ for the positions, and $|\mathbf{x(3:4)}| \leq 1m/s$ for the velocities. The control input bounds are $|\mathbf{u}| \leq 0.25 m/s^2$. The MILP solver attempts to minimize the cost function mentioned in \ref{eq15}. The very large number \textit{M} is initialised as $10^{3}$ and the minimum desired robustness as $\rho^\phi_{min} = 0.2$. The time range in seconds $[a,b]$ is given as [15,20] for the \textit{Go-To-Goal} example and ([15,20], [25,30]) for the two waypoints in the \textit{Multi-Waypoint} example. The simulation results are as shown in fig: (\ref{fig2}) and (\ref{fig3})

The lighter boxed areas around the {\color{Green} goal} and \textbf{obstacle} regions denote the boundary for minimum robustness($\rho^\phi_{min}$) desired around the obstacle. The {\color{Red} AV} is shown over the trajectory as it progresses toward the goal region. We observed the following results for the examples:
\begin{itemize}
    \item For the \textit{Go-To-Goal} example, the system was completely safe with positive robustness throughout the simulation period. The robustness boundary areas for the lower obstacle's two corners are violated but still fall within $0 <\rho^\phi(t) < 0.2$.The solver takes \textbf{3.5} seconds to calculate a feasible trajectory after which the STL specification asks for the AV to reach the goal within \textit{20} secs.
    \item For the \textit{Multi-Waypoint} example, the system is again completely safe with minor discrepancies. The target time range to reach each waypoint was also satisfied, as the velocities were observable and controllable states that could vary within the provided range. The solver takes \textit{4.7} seconds to calculate a feasible trajectory. This is a bit longer than the first example, given the complexity of the problem at hand.
    \item There were a few infeasible solutions that we saw initially during the simulations. They were primarily due to providing the incorrect time ranges for \{\ref{ex4}\}. This was because it was difficult to gauge the exact time range in which the AV would reach the waypoints. A workaround is it includes larger time frames such that $``ev_{[a,b]}"$ operator has a larger frame to compute within the control input bounds.
\end{itemize}
\section{Conclusion}\label{s6}
In this paper, we focused on developing a signal temporal logic based 
constraints approach to find safe and optimal solutions for autonomous vehicle navigation tasks. We use a model predictive control approach to apply the optimal control inputs calculated using the mixed integer linear solver. The STL constraints are user inputs provided to the model, first parsed to identify important information such as goal coordinates, obstacle regions, and time ranges. Once we acquire this information, along with the state and control bounds, we formulate an MILP problem that is solved recursively subject to satisfaction and maintenance of a minimum positive robustness value throughout the simulation. The model, constraints, and cost function are all linear in length, reducing calculation time as we are required to calculate smaller MILPs for most of the problem. Considering that we apply the safety constraints only at critical locations along the trajectory, we are not always guaranteed an optimal solution.
 We apply the approach to an AV model particularly focusing on safety for an autonomous navigation task. Two navigation task-focused numerical examples are solved, where feasible trajectories are generated. The computed optimal solutions are applied to navigate the AV safely to the waypoints. Future work for this approach demands expanding the application to non-linear systems to conduct real-time experiments on smaller vehicles in a controlled environment. We are also planning on making the model robust to disturbances as our application space is eventually for off-road 3D environments with higher fidelity models and environment. The STL parser would also be upgraded to work with multiple constraints without causing LIE explosion for the MILP solver. We aim to create efficient models for real-time applications on off-road vehicles.

\bibliographystyle{ieeetr} 
\bibliography{sampleBib}

\section{Contact Information}
 Aditya Parameshwaran\\
 Graduate Research Assistant, Department of Mechanical Engineering, Clemson University \\ 
 e-mail: aparame@clemson.edu
 
 Yue Wang\\
 Professor, Department of Mechanical Engineering, Clemson University \\ 
 e-mail: yue6@g.clemson.edu

\section{Acknowledgments}
This work was supported by the Simulation Based Reliability and Safety Program for modeling and simulation of military ground vehicle systems, under the technical services contract No. W56HZV-17-C-0095 with the U.S. Army DEVCOM Ground Vehicle Systems Center (GVSC).
\\ Distribution A. Approved for public release; distribution unlimited. (OPSEC 7039)

\clearpage
\end{document}